# Periodicity Detection of Outlier Sequences Using Constraint Based Pattern Tree with MAD

Ms. Archana N.
ME Student, Department of Computer Engineering
D.Y. Patil College of Engineering, Akurdi
Savitribai Phule Pune University, Pune, India
achuarc.n@gmail.com

Ms. S. S. Pawar
Asst. Prof, Department of Computer Engineering
D.Y. Patil College of Engineering, Akurdi
Savitribai Phule Pune University, Pune, India
psoudamini@yahoo.co.in

*Abstract*— **Patterns that appear rarely or unusually in the data can be defined as outlier patterns. The basic idea behind detecting outlier patterns is comparison of their relative frequencies with frequent patterns. Their frequencies of appearance are less and thus have lesser support in the data. Detecting outlier patterns is an important data mining task which will reveal some interesting facts. The search for periodicity of patterns gives the behavior of these patterns across time as to when they repeat likely. This in turn helps in prediction of events. These patterns are found in Time series-data, social networks etc. In this paper, an algorithm for periodic outlier pattern detection is proposed with the usage of a Constraint Based FP (Frequent Pattern)-tree as the underlying data structure for time series data. The growth of the tree is limited by using level and monotonic constraints. The protein sequence of bacteria named E.Coli is collected and periodic outlier patterns in the sequence are identified. Further the enhancement of results is obtained by finding the Median Absolute Deviation (MAD) in defining candidate outlier patterns. The comparative results between STNR-out (Suffix Tree Noise Resilient for Outlier Detection) and proposed algorithm are illustrated. The results show the effectiveness and applicability of the proposed algorithm.**

## I. Introduction

A time series is a sequence of observations of well-defined data items obtained through repeated measurements over time**.** Level of employment measured every month can be considered as an example of time series. A time series is usually discretized before analysis. "Pattern mining" finds existing patterns in a given data collection. There are several categories of patterns that can be mined from the data such as frequent pattern mining, sequential pattern mining etc. But finding only frequent patterns may not reveal the behavior of data. Researches on analysis of data show that detection of outlier patterns might be more important in many sequences than regular, frequent patterns. The periodicity of these outlier patterns reveals interesting facts. For example it would help in prediction of drought and flood, stock market price prediction, earthquake prediction and so on. There are many techniques to find local or global outliers and frequent patterns in the data but detecting outlier or surprising patterns are different from detection of other patterns. Periodicity detection of outlier patterns is an area of research yet to be explored in detail.

Researches on periodic pattern mining have always come up with algorithms which give more significance to patterns that have higher support within the analyzed sequence. For example consider power consumption patterns of a housing community. Periodic pattern mining focuses on how much power is being consumed frequently by houses in that particular community. But there might be a large number of power consumption being reported at sudden point of time. This abnormal activity is an outlier pattern which has lesser support amongst all reports. Analyzing this outlier pattern reveals its occurrence weekly. This is called as periodicity of outlier patterns.

Periodic Outlier Patterns can also be found in ECG pulse rate with unusual reports at certain point of time, weather data where temperature can show surprising measurements, stock market data where share rate change all of a sudden, retail market data where customer behavior might change in certain periods of time, medical data about some severe epidemic over time etc. All these are examples of time series data where outlier pattern can occur repeatedly over a period of time. Fig. 1 shows anomalous pattern in heart beat rate over time. The problem statement of this paper is to detect how periodically this pattern occurs. Finding periodicity of outlier patterns could reveal important observations about the behavior and future trends of the data. Periodic outlier pattern detection using constraint based periodicity mining uses a Tree-like data structure called as consensus tree.

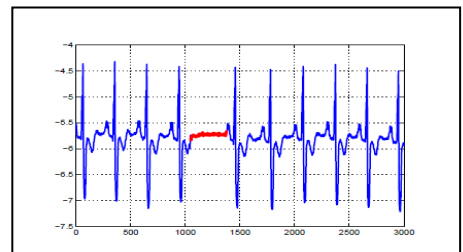

Figure 1.  Anomalous pattern in heart beat rate [1].

The growth of this FP tree is restricted by providing additional mining constraints. Hence it is space and time efficient compared to existing algorithms. This tree was used by Dr. Ramachandra V.Pujeri, G.M. Karthik [9] for finding symbol, partial and full periodicity in time series. In this paper it is further extended where the tree is generated to represent the frequent and outlier patterns. The tree is further annotated to mine periodicity of outlier patterns.

Finally, the contribution described in this paper can be given as follows: 1) the use of a FP consensus tree-based time efficient algorithm for unusual or outlier periodic patterns; the tree growth is limited by using certain constraints 2) the usage of Median Absolute Deviation for enhancement of results; 3) detecting periodic outlier patterns in protein sequence of an E.Coli bacterial data set collected from UCI machine learning repository; 4) comparative results between existing and proposed algorithm. Protein sequence can be considered as a time series data.

## II. RELATED WORK

There are several kinds of periodic patterns that can be present in a collection of data. Sirisha and Padma Raju [2] have classified the periodic patterns in 5 different ways. Full periodic pattern is a pattern where every position in the pattern exhibits periodicity. Periodic patterns in which one or more elements do not exhibit the periodicity are called partial periodic patterns. In the input sequence {a}{b}{c}{b}{c}{b}{c}{a}{c}{d}, {b}{c} is a full periodic pattern with period 2. {a}{*}{c} is a partial periodic pattern with period 3 in the sequence {a}{b}{c}{a}{d}{d}{a}{c}{c}. A pattern X is said to satisfy perfect periodicity in sequence S with period p if starting from the first occurrence of X until the end of S, every next occurrence of X exists p positions away from the current occurrence of X. {a}{b}{*} is perfect periodic pattern with period 3 in the sequence {a}{b}{d}{a}{b}{v}{a}{b}{f}{a}{b}{c}. A pattern which occurs periodically without any misalignment is called as synchronous periodic pattern. A sequence is said to have symbol periodicity if at least one symbol is repeated periodically. A pattern consisting of more than one symbol repeating with same periodicity in a sequence leads to sequence periodicity. If the whole sequence can be mostly represented as a repetition of a pattern or segment then that type of periodicity is called segment or segment-cycle periodicity. C. Sheng, W. Hsu, and M. L. Lee [3] detect sequence periodicity. For example in the sequence {x}{y}{y}{x}{u}{i}{x}{t}{r}{x}{k}{l}, symbol{x} is periodic with periodicity 3.

There are several existing algorithms for detecting periodicity in time series databases which can be classified based on the type of periodicity they detect; for example, some detect symbol periodicity, while some others partial periodicity, segment periodicity and so on. M. Elfeky, W. Aref, and A. Elmagarmid [4] have proposed a time warping algorithm, named WARP(Time Warping for Periodicity Detection), for periodicity detection in the presence of noise. The algorithm shifts the time series by p positions and compares the original time series to the shifted version to detect segment periodicity. But it failed to find symbol or sequence periodicity.

J. Han, G. Dong, and Y. Yin [5] have detected partial periodic patterns (ParPer) by mining association rule i.e. a pattern is a frequent partial pattern in the time series if its confidence is larger than or equal to a threshold min_conf. The efficient mining of partial periodic patterns is performed by authors for a *single period* as well as for a *set of periods*.

M. Elfeky, W. Aref, and A. Elmagarmid [6] proposed convolution based periodicity detection algorithm (CONV). The authors have detected two types of periodicity; segment periodicity and symbol periodicity. The idea behind the algorithm for segment periodicity detection was to use the concept of convolution in order to shift and compare the time series for all possible values of the period.

F. Rasheed and R. Alhajj [7] proposed Suffix tree Noise Resilient algorithm (STNR) for detecting all types of periodicity. In this approach the numerical data is discretized into a data of numerals. A suffix tree representation for the numerical data is developed next. The tree is annotated to give the occurrence vector of a substring. The difference in the occurrence positions gives the periodicity.

Jisha Krishnan and Chitharanjan K have studied different periodicity detection algorithms and done comparison among four above stated algorithms [8]. CONV gives best time performance in comparison to WARP, ParPer and STNR.

Dr. Ramachandra V.Pujeri, G.M. Karthik [9] have proposed a Constraint Based Periodicity Mining approach where the periodicity is mined for frequent patterns based on certain constraints on the growth of a FP (Frequent Pattern) Tree. Huang and Chang [10] presented their algorithm for finding asynchronous periodic patterns, where the periodic occurrences can be shifted in an allowable range within the time axis.

Another way of classifying the existing work for periodicity detection on time-series analysis is where the first category includes algorithms that require the user to specify the period (or the maximum period) and then look only for patterns occurring with the specified period (or up to the maximum period), and the second class are algorithms which look for all possible periods in the time series. CBPM (Constraint Based Periodicity Mining) looks for all possible periods starting from all possible positions.

In [11] F. Rasheed and R. Alhajj have extended STNR further for periodic outlier pattern detection. They compare their work to give better performance with the InfoMiner algorithm [12]. E. Keogh, J. Lin, and A. Fu [13] identify subsequence as outliers on 82 different time-series datasets using symbolic aggregate approximation.

B. Janani and S. Rajkumar [14] have used the framework proposed in [11] for periodicity detection of outlier pattern for weather forecasting system.

Archana and S.S. Pawar [15] give a brief overview on different techniques for outlier pattern detection in time-series data.

Leys et al. [16] show that usage of MAD is better than mean in detecting outliers from a set of data.

### III. PROBLEM DEFINTION

The pattern $X = cd$ with period $p = 7$ is a better candidate for outlier pattern in the sequence S as given below – with pattern at top and position index in the second row.

| c | b | d | c | b | d | c | **c** | **d** | c | b | d | c | b | **c** | **d** |
|---|---|---|---|---|---|---|---|---|---|---|---|---|---|---|---|
| 0 | 1 | 2 | 3 | 4 | 5 | 6 | 7 | 8 | 9 | 0 | 1 | 2 | 3 | 4 | 5 |
| c | b | d | c | b | **c** | **d** | c | b | d | c | b | d | c | b | d |
| 6 | 7 | 8 | 9 | 0 | 1 | 2 | 3 | 4 | 5 | 6 | 7 | 8 | 9 | 0 | 1 |

The pattern $X = cd$ with period $p = 2$ is not a good candidate for outlier pattern in the sequence S' given below.

| c | b | d | c | b | d | **c** | **d** | **c** | **d** | **c** | **d** | c | b | d |
|---|---|---|---|---|---|---|---|---|---|---|---|---|---|---|
| 0 | 1 | 2 | 3 | 4 | 5 | 6 | 7 | 8 | 9 | 0 | 1 | 2 | 3 | 4 | 5 |
| c | b | d | c | b | d | c | b | d | c | b | d | c | b | d | c |
| 6 | 7 | 8 | 9 | 0 | 1 | 2 | 3 | 4 | 5 | 6 | 7 | 8 | 9 | 0 | 1 |

A less frequent pattern with larger coverage area (having repetitions in larger subsection of the sequence) is more interesting than those with smaller coverage area (repeating in smaller subsection of the sequence) [11].

Periodicity Detection involves finding confidence of a pattern in a given series. The basic terms of calculating confidence and surprise of a pattern which is borrowed from [11] is explained below with an example.

Consider the series S given below:

| **x** | **y** | a | e | **x** | **y** | b | d | **x** | **y** | z | d | **x** | **y** |
|---|---|---|---|---|---|---|---|---|---|---|---|---|---|
| 0 | 1 | 2 | 3 | 4 | 5 | 6 | 7 | 8 | 9 | 0 | 1 | 2 | 3 |
| b | d | **x** | **y** | z | d | x | b | y | y | **x** | **y** | z | y |
| 4 | 5 | 6 | 7 | 8 | 9 | 0 | 1 | 2 | 3 | 4 | 5 | 6 | 7 |

Pattern $X = xy$ is repeating in S with period: $p = 4$, starting at position $i_{st} = 0$, ending at position $i_{end} = 25$, with pattern length: $|X| = 2$. The maximum number of repetition is calculated as (1).

$$f_{max} = \frac{i_{end} + 1 - |X| - i_{st}}{p} + 1 \qquad (1)$$

Thus $f_{max} = (25+1-2-0) / 4 +1 = 24/4 + 1 = 7$.
Thus conf (xy, 0, 25, 4) given by $f/ f_{max} = 6/7$.

A periodic pattern $X$ is said to be *frequent* if its confidence is greater than or equal to a user-defined threshold $conf_{min}$, as in (2).

$$conf(X, i_{st}, i_{end}, p) > conf_{min} \qquad (2)$$

Two other parameters required are *minSegLen* to specify the minimum length of valid periodic segment. Another parameter that is used in the optimization is called $d_{max}$, which is the maximum allowable distance between any two periodic occurrences of a pattern. If $f(X)$ represents the frequency (repetition count) of the pattern $X$, and $segLen(X)$ represents the segment length of the repetitions of $X$, then $X$ is *candidate outlier pattern* if (3) is satisfied.

$$f(X) < MAD(f(X_i)) \text{ AND } segLen(X) > minSegLen; \qquad (3)$$

$\forall$ i such that $|X_i| = |X|$ where $MAD (f(X_i))$ is the median absolute deviation of the frequency of all patterns of length exactly the same as that of pattern $X$. Dhwani Dave, TanviVarma [17] have used MAD to replace mean in the framework proposed in [11] for periodicity detection of outlier patterns in oil prices to obtain better results.

The measure of *surprise* of a pattern $X$ is defined as one minus the ratio of the frequency of $X$ over the average frequency of all patterns with same length as $X$ given in (4).

$$surprise(X) = 1 - \frac{f(X)}{\mu(f(x_i))}; \forall i \qquad (4)$$

A candidate outlier pattern $X$ is an *outlier periodic pattern* iff (5) is satisfied.

$$surprise(X) > surprise_{min} \text{ AND } conf(X, i_{st}, i_{end}, p) > conf_{min}$$
$$\text{———(5)}$$

where $conf(X, i_{st}, i_{end}, p)$ is the confidence of pattern $X$ repeating with period $p$ within the segment starting and ending at $i_{st}$ and $i_{end}$, respectively; $conf_{min}$ and $surprise_{min}$ are the minimum confidence and minimum surprise values provided by the user respectively. The surprise criteria in the last definition can be avoided with the usage of MAD.

### IV. OUTLIER PATTERN IDENTIFICATION

The proposed algorithm for periodic outlier pattern detection uses Constraint Based Periodicity Mining. Hence the name for the algorithm goes as CBPM-out; where 'out' stands for outlier detection. This algorithm takes the input series data and identifies periodic outlier patterns in it. Fig. 2 gives the flowchart for Preprocessing Module in the system. Fig. 3 gives the flowchart for Periodic Outlier Pattern detection Module using Constraint Pattern Tree Generation.

*A. Overview of CBPM-out*

CBPM-out involves the following steps:

**1) *Discretization of Data*** - The discretization process transforms the time series into a series consisting of a finite set of symbols.

***Constraint Based FP Tree Construction*** – This step involves the construction of a consensus tree referred to as FP (Frequent Pattern) Tree. In general, each frequent itemset is represented as a path of a tree, from root to some leaf node. The consensus tree growth is pre-pruned based on the constraints. CBPM (Constraint Based Periodicity Mining) technique uses level and monotonic constraints so that the growth of the tree is restrained. Nodes with confidence value as conf (b) = (N- (sup (b))/ (N-q) < min_conf will be pruned; it is used as an monotonic constraint. Here N is the number of the input sequence and q is a positive integer $1 \leq q \leq N$. min_conf is a value specified by user.

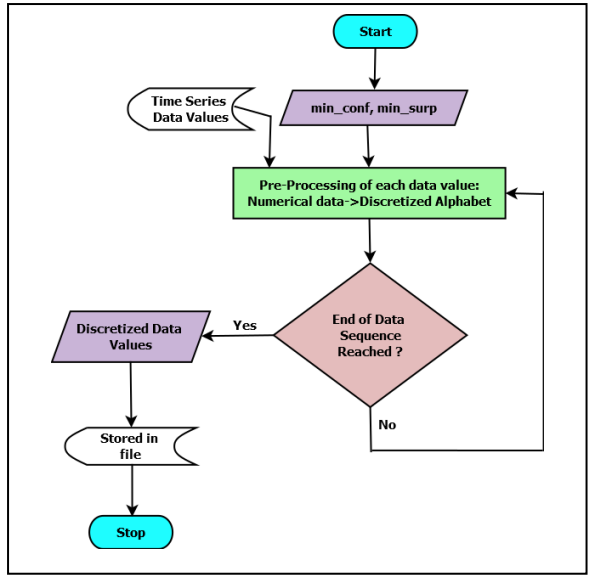

Figure 2. Flowchart for Preprocessing of data

A node in the consensus tree will branch out only if it possesses a (level of the tree at that node) ≤ q; where q represents the 0.5 of the length of the sequence. This is used as a level constraint [9].

Due to this optimization strategy of constraints there is no redundant comparison further. For an example of monotonic constraint consider the input given in Fig. 4. The pattern 'cdc' will not be represented in the tree because it is not frequent.

2) *FP Tree Annotation* – The FP tree is annotated to maintain position pointers of a pattern. Each node holds a Position vector for the pattern it is representing. Fig. 4 depicts the consensus annotated FP Tree for the series given in square box on top of the same figure.

3) *Creation of Position Vector* – The position pointers in each node leads is nothing but the position vector. The Pattern Frequency Table(PFT) will look like Table I(for the input data here).

TABLE I. PFT

| Pattern Length | Mean | MAD |
|---|---|---|
| 1 | 55.5 | 23.72 |
| 2 | 22.13 | 9 |
| 3 | 18 | 5.93 |
| 4 | 11 | 4.4478 |

4) *Detect Candidate Outlier Patterns* – The patterns likely to be outlier patterns are detected. These are patterns which are less frequent than the patterns of same length using MAD(Mean Absolute Deviation).

**Advantage of using MAD in definition of candidate outlier pattern:** Consider a set of data of 13 observations as given below which is giving the frequency of patterns of a given length: *1, 2, 4, 5, 7, 8, 200, 250, 270, 1100, 1105, 1200, 1500.*

The mean and standard deviation of the above data would be 434.76 and 543.441 respectively. Using the criterion of mean, data less than 3*Mean i.e. data less than 1304.28 will be accepted as normal data. Using the criterion of standard deviation, data less than 3*S.D i.e. data less than 1630.32 will be accepted as candidate outliers because they are less frequent when compared to other frequent patterns.

We would not want pattern with frequency of above 1000 to be detected as outliers because it is not so surprising. Using MAD solves the above problem.

Median of the above data will be 7$^{th}$ value which is 200. MAD is calculated using the formula in (6) [16].

$$MAD = b\ M_i (|\ x_i - M_j\ (xj)\ |) \quad (6)$$

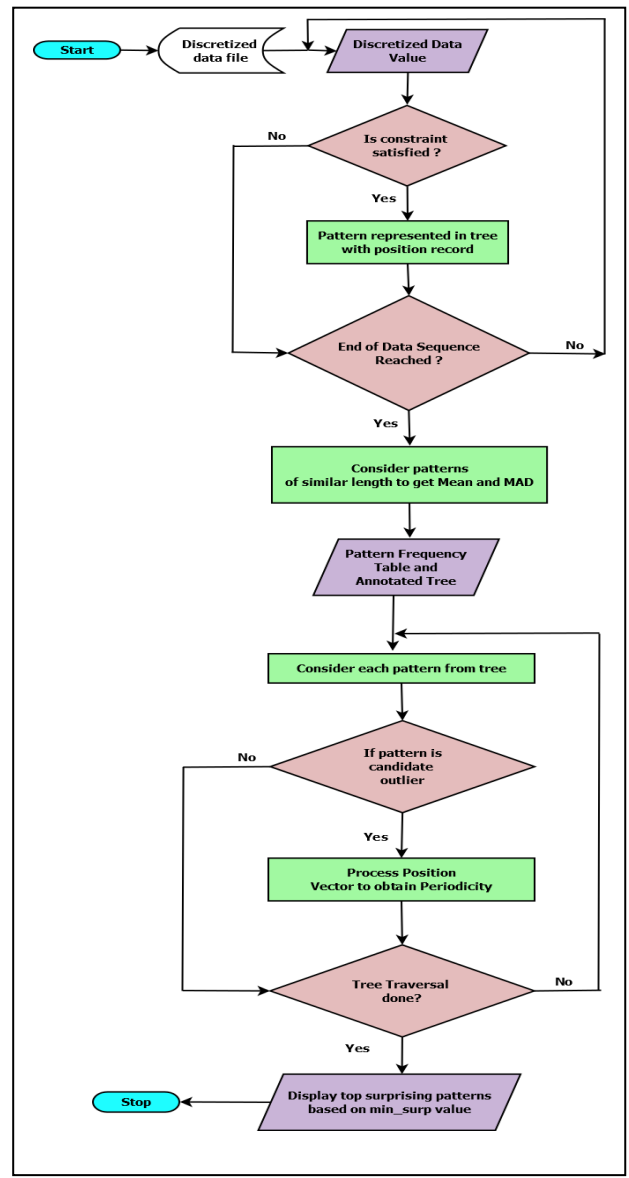

Figure 3. Flowchart for Periodic Outlier Pattern detection

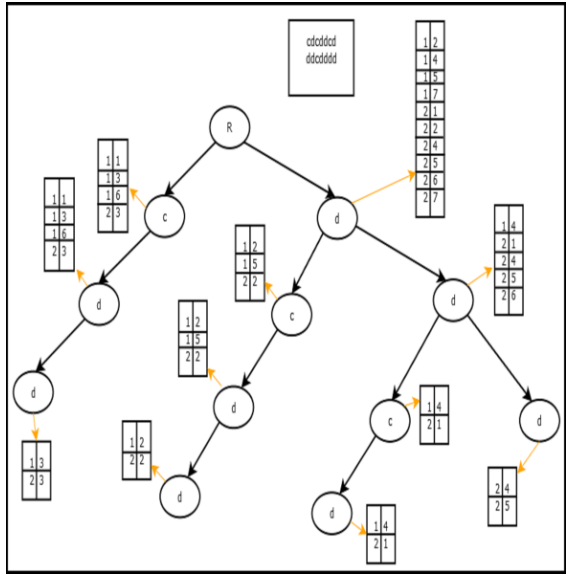

Figure 4. Constraint Based FP Tree

Calculating MAD involves the following steps. The median is subtracted from each observation taking the absolute values. |1-200|, |2-200|, |4-200|, |5-200|, |7-200|, |8-200|, |200-200|, |250-200|, |270-200|, |1100-200|, |1105-200|, |1200-200|, |1500-200|.

The series now arranged in ascending order becomes 0, 50, 70, 192, 193, 195, 196, 198, 199, 900, 905, 1000, 1300. Now the median becomes $7^{th}$ value which is 196. Usually $b = 1.4826$ is linked to the assumption of the normality of data. Hence multiplying 196 with 1.4826 gives a MAD = 290.58. As per research a threshold of 3 will be used for the decision criterion. Hence data less than 3*MAD i.e. data less than 871.76 will be found as outliers.

 5) *Find Periodicity of outlier pattern* – The difference between the two successive positions of the position vector is the periodicity of the pattern.

 6) *Find periodic outlier patterns* – *T*op surprising patterns would be displayed based on min_surprise given by user.

*B. CBPM-out Algorithm*

It includes two main phases.

 **1)** Construct FP/consensus tree for the input data sequence with min_conf and min_sup specified by user; pseudocode shown in Algorithm 1.

 **2)** Periodicity of Outlier patterns are mined from the tree with min_surp specified by user; pseudocode shown in Algorithm 2.

**Algorithm 1:** FP-tree construction

- **Input:** Protein Sequence data, min_conf
- **Output:** Annotated FP Tree
1. Initialise the tree with root node.
2. **for** each symbol of input string **do**
3. **loop(1):**
4. **if** (!(symbol exist)) create new node
5. update position vector.
6. **for** each consecutive symbol **do**
7. **loop(2):**
8. **if** pattern already represented in the tree
  update the position vector.
 **else**
 8.1 **if** monotonic and level constraints satisfied
   Create new node to represent the pattern.
   Annotate tree to hold the position of the pattern.
 8.2 **else**
   Prune the growth. Go to Step 2.
9. e**nd loop(2)**
10. **end loop (1)**

**Algorithm 2:** Periodicity Mining

- **Input:** Position Vector and time tolerance t
- **Output:** Periodicity of Outlier Patterns
1. **for** patterns of same length in the tree **do**
2. **loop (1):**
3. Detect candidate outlier patterns from tree using MAD.
4. **end loop (1)**
5. **for** each candidate outlier pattern **do**
6. **loop (2):**
7. Process position vector to detect periodicity considering the time tolerance window.
8. **end loop (2)**

Finally top surprising patterns with periodicity displayed based on min_surp specified by user.

V. EXPERIMENTAL EVALUATION

The algorithm is experimented on E.Coli protein sequence. Details and application about E.Coli can be found in [18], [19]. The E.Coli data set has been downloaded from the link https://archive.ics.uci.edu/ml/datasets/Ecoli.

*A. Accuracy*

The existing algorithm i.e. STNR-out [11] and proposed algorithm i.e. CBPM-out are compared by implementing them using E.Coli protein sequence. The existing algorithm uses Suffix Tree for detection of periodic outlier patterns [11]. The proposed algorithm is both time and space efficient due to use of constraints. The definition of the candidate outlier pattern is also improved in this proposed algorithm. The parameters of the surprising patterns obtained when experimenting using both the algorithms are shown below in Table II and III respectively. Suffix tree with mean gives less accurate results when compared to consensus based-FP Tree with MAD i.e. more surprising patterns obtained in the proposed system.

TABLE II. DETAILS ABOUT SURPRISING PATTERNS OBTAINED USING STNR-OUT

| Count | Period | Pattern | Start Pos | End Pos | Conf | Surp |
|---|---|---|---|---|---|---|
| 2 | 97 | D | 212 | 309 | 1 | 0.9639 |
| 2 | 5 | F | 324 | 329 | 1 | 0.9639 |
| 2 | 10 | C | 112 | 122 | 1 | 0.9639 |
| 4 | 5 | E | 308 | 328 | 0.8 | 0.9279 |
| 5 | 16 | B | 16 | 80 | 1 | 0.9099 |
| 2 | 56 | CA | 95 | 152 | 1 | 0.9096 |
| 2 | 20 | DF | 309 | 329 | 1 | 0.9096 |
| 4 | 30 | CD | 63 | 154 | 1 | 0.819 |
| 2 | 14 | BAEH | 413 | 428 | 1 | 0.8181 |
| 3 | 10 | BAFG | 183 | 206 | 1 | 0.727 |
| 11 | 16 | GH | 1 | 178 | 0.8 | 0.5029 |
| 11 | 16 | HE | 2 | 179 | 0.8 | 0.5029 |
| 6 | 16 | DDEH | 231 | 304 | 1 | 0.4545 |
| 6 | 14 | DFEH | 315 | 388 | 1 | 0.4545 |
| 11 | 16 | GHE | 1 | 179 | 0.8 | 0.3888 |
| 16 | 15 | EH | 219 | 430 | 1 | 0.2769 |

TABLE III. DETAILS ABOUT SURPRISING PATTERNS OBTAINED USING CBPM-OUT

| Count | Period | Pattern | Start Pos | End Pos | Conf | Surp |
|---|---|---|---|---|---|---|
| 2 | 97 | D | 212 | 309 | 1 | 0.9639 |
| 2 | 5 | F | 324 | 329 | 1 | 0.9639 |
| 2 | 10 | C | 112 | 122 | 1 | 0.9639 |
| 4 | 5 | E | 308 | 328 | 0.8 | 0.9279 |
| 5 | 16 | B | 16 | 80 | 1 | 0.9099 |
| 2 | 56 | CA | 95 | 152 | 1 | 0.9096 |
| 2 | 20 | DF | 309 | 329 | 1 | 0.9096 |
| 4 | 30 | CD | 63 | 154 | 1 | 0.819 |
| 2 | 14 | BAEH | 413 | 428 | 1 | 0.8181 |
| 3 | 10 | BAFG | 183 | 206 | 1 | 0.727 |

*B. Time Performance*

Fig. 5 and Fig. 6 depict the time based graph comparison between STNR-out proposed by Rasheed et al.[11] and proposed approach i.e. CBPM-out in this paper.

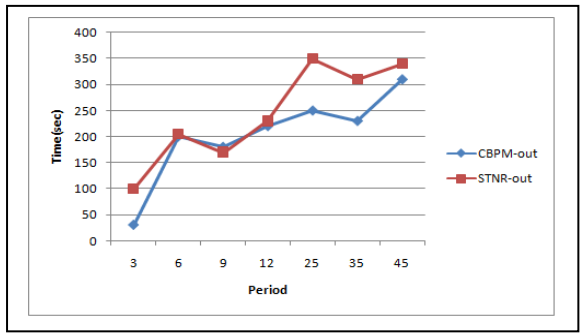

Figure 5. Time performance of CBPM-out with STNR-out with increasing period size

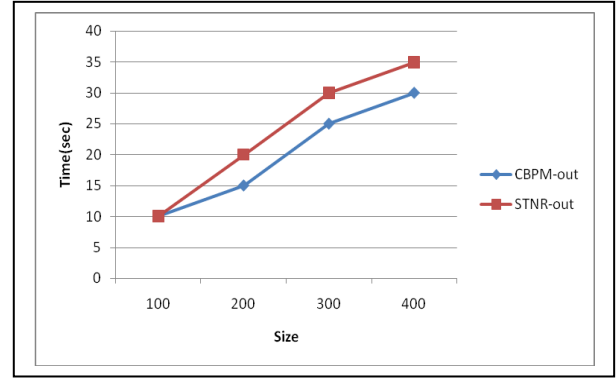

Figure 6. Time performance of CBPM-out with STNR-out with increasing series length

Fig. 5 shows that with varying period size STNR-out consumes more time than CBPM-out. Fig. 6 shows that as data size increases; time taken by CBPM-out is relatively lesser than STNR-out. This behavior is understandable because the usage of constraints in the growth of the tree reduces time consumption in the entire framework. Another reason for reduction in time in proposed algorithm is pruning of unnecessary pattern generation.

VI. CONCLUSION

Detecting Periodic outlier patterns is a kind of pattern mining which helps in prediction and forecasting of the events. It might be more important than periodicity of frequent patterns. The protein sequence can be manipulated easily for E.Coli bacteria and helpful in discovery of vaccines further. For periodicity mining Constraint based Tree has been used. Surprising or unusual pattern takes into account the relative frequency of a pattern with patterns of similar length. Time tolerance window enables to work with noisy series data. The definition of surprising patterns is also improved using Median Absolute Deviation. The proposed algorithm consumes lesser time due to usage of constraints.

VII. FUTURE WORK

The system can be further improved for fuzzy data dredging and online periodicity detection. Further periodicity detection can be done in fuzzy time series data input.

ACKNOWLEDGMENT

The authors would like to thank the publishers, researchers for making their resources available and teachers for their guidance. We also thank the Savitribai Phule Pune University and college authority for providing the required infrastructure and support. Finally we would like to extend a heartfelt gratitude to friends and family members.